\documentclass[]{spie}  

\usepackage{amsmath,amsfonts,amssymb}
\usepackage{graphicx}
\usepackage[colorlinks=true, allcolors=blue]{hyperref}

\title{A comparison of solar and stellar coronagraphs 
that make use of external occulters }

\author{Claude Aime}
\author{Céline Theys}
\author{Simon Prunet}
\author{André Ferrari}
\affil{Université Côte d'Azur, Observatoire de la Côte d'Azur, CNRS, Nice, France}

\authorinfo{claude.aime@univ-cotedazur.fr}

\begin{document} 
\maketitle

\begin{abstract}
Solar and stellar externally occulted coronagraphs  share similar concepts, but are actually very different because of geometric characteristics. Solar occulters were first developed with a simple geometric model of diffraction perpendicular to the occulter edges. We apply this mere approach to starshades, and  introduce a simple shifted circular  integral of the occulter which allows  to illustrate the influence of the number of petals on the extent of the deep central dark zone. We illustrate the reasons for the presence of an internal coronagraph in the solar case and its absence in the exoplanet case.
\end{abstract}


\section{INTRODUCTION}
\label{sec:intro}  

\begin{figure} [ht]
   \begin{center}
   \begin{tabular}{c} 
   \includegraphics[height=5cm]{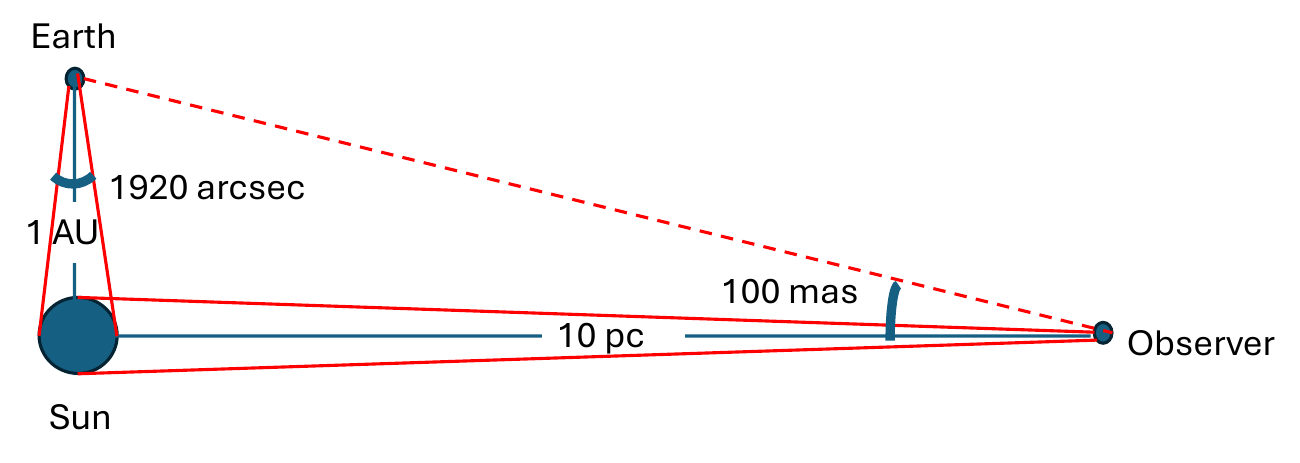}
   \end{tabular}
   \end{center}
   \caption[example] 
   { \label{schematic} 
Illustration of characteristic dimensions for solar and stellar coronagraphs. Distances are from 1 AU to 10 pc. }
   \end{figure} 

Astronomical objectives for solar and stellar coronagraphs are  the study of the solar corona and the detection of exoplanets. We consider here coronagraphs using external occulters on the model of the occultation of the Sun by the Moon.
These  coronagraphs share similar concepts but become different because of geometric characteristics. In round numbers, distances between the objects to be occulted by these coronagraphs are as 1 astronomical unit (AU) for the solar corona vs  10 parsecs or more for exoplanets, a factor of  2 million or so.  A very schematic comparative representation of the observational parameters is given in Fig.~\ref{schematic} and a list of experimental parameters is given in Table \ref{Comparison}. From a  near-Earth orbit, the Sun is seen at an angle of about half a degree. We seek to see the solar corona as close as possible to the solar disk, and the external occulter is just greater than   the apparent diameter of the solar photosphere. The corona is observed over a few solar radii. For exoplanets, considering an extra-solar system at 10 parsecs, the angle between the exo-Sun  and the exo-Earth  is  100 milliarcsec, and the stellar diameter would be seen at an angle of 1 mas. Starshades, with all the experimental constraints, do not have the possibility to just obscure the 1 mas star, and consider much larger obstruction angles,   50 to 100 times as large, just to allow the detection of  an exo-Earth. A heuristic and historical presentation of the ideas leading to the diversity of external solar occulters is made  in Section \ref{Sunshade}. A similar presentation for exoplanets is given  in Section \ref{Starshade}, which gives the opportunity to introduce a function which can be considered as a shifted apodization profile. An empirical justification of the usefulness of an internal coronagraph in the solar case and its low efficiency in the case of exoplanets is given in Section \ref{Lyot}.

\begin{table}[ht]
\caption{Comparison between solar and stellar coronagraphs. In addition to the orders of magnitude between  dimensions, a main point is  that the solar disk is a huge extended object  versus the almost point source star. Values are for a typical extra-solar system seen at 10 parsecs. } 
\label{Comparison}
\begin{center}       
\begin{tabular}{|l|l|l|} 
\hline
\rule[-1ex]{0pt}{3.5ex}  Science object & Solar Corona & Exoplanets  \\
\hline
\rule[-1ex]{0pt}{3.5ex}  Occulter diameters & A few cm. to 1.5 m. & 10 m. to 100 m.   \\
\hline
\rule[-1ex]{0pt}{3.5ex}  Occulter shapes & Single or multiple discs, Serrated, Multithreaded ... & Petal-shaped   \\
\hline
\rule[-1ex]{0pt}{3.5ex}  Occulter telescope distance & A few cm., up  to 150 m   & 10 to 100 $\times 10^{6}$ m.\\
\hline
\rule[-1ex]{0pt}{3.5ex}  Telescope diameter & A few cm., up to 5cm & 1 m. to 6.5 m.   \\
\hline
\rule[-1ex]{0pt}{3.5ex}  Straylight origin & Solar photosphere & The star   \\
\hline
\rule[-1ex]{0pt}{3.5ex}  Type of source & Very extended (1920 arcsecs) & Point-like (1 mas)  \\
\hline
\rule[-1ex]{0pt}{3.5ex}  Linear res. pixels in the source & 1000 & 0.01  \\
\hline
\rule[-1ex]{0pt}{3.5ex}  IWA/$\Phi$ source & 1 - 1.03 & 30 - 70  \\
\hline
\rule[-1ex]{0pt}{3.5ex}  Field of view & A few degrees & 1 arcsec  \\
\hline
\rule[-1ex]{0pt}{3.5ex}  Required rejection & $10^{-6} - 10^{-9}$ & $10^{-6} - 10^{-12}$ \\
\hline
\rule[-1ex]{0pt}{3.5ex}  Internal coronagraph & Lyot type & No \\
\hline
\rule[-1ex]{0pt}{3.5ex}  Fresnel number & $10^{2} - 10^{4}$  & 1 - 10\\
\hline
\end{tabular}
\end{center}
\end{table} 

\section{SOLAR EXTERNAL OCCULTERS }
\label{Sunshade} 
The use of external occulters to shade the bright solar photosphere to observe the faint solar corona without natural eclipses has a long history.
Solar astronomers considered  first that  Lyot's coronagraph \cite{Lyot1933}, with the rather low quality of optical  systems of their time,  cannot be improved enough to observe   the outer corona \cite{NewBoh1963, Newkirk1965}.  Then, with reference to Evans's photometer \cite{Evans1948}, an external occulter is  set in front of the Lyot coronagraph in almost all space-born coronagraphs, as described by  Koutchmy \cite{Koutchmy1988}. In overall, the effects of the external occulter provide the same rejection rate as that of the Lyot coronagraph, and these effects multiply. 

Observers were very inventive. 
In 1963, Tousey \cite{Tousey1965}, used a serrated-edge external-occulter and obtained the first  observations of the outer solar corona in the absence of an eclipse. Purcell and Koomen \cite{Purcell1962} proposed to use a  system of multiple circular disks. A decisive step was made with the coronagraph  SOHO/LASCO \cite{Brueckner1995},  in operation since 1995.
In their experimental study of new generations of external occulters for LASCO-C2, they { \cite{Bout:00}} found that conic occulters gave performances superior to   multiple discs occulters.
\begin{figure} [ht]
   \begin{center}
   \begin{tabular}{ccc} 
   \includegraphics[height=4cm]{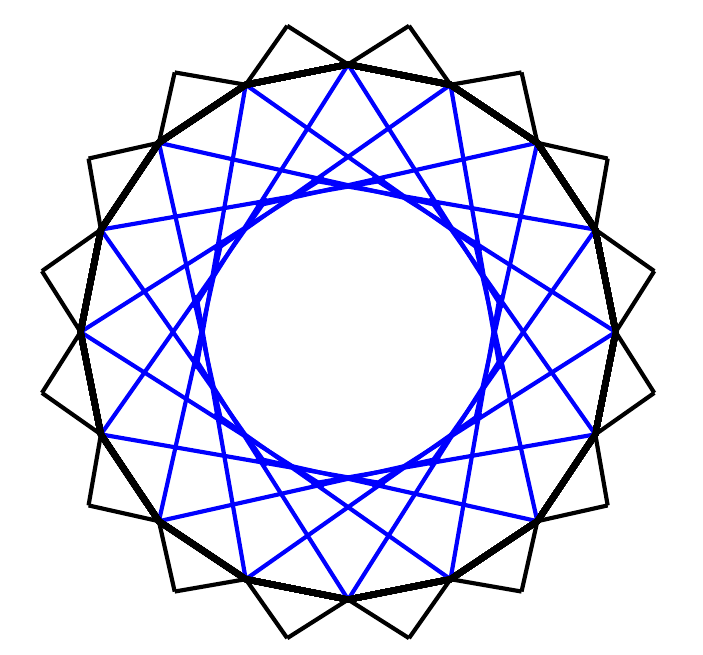} & \includegraphics[height=4cm]{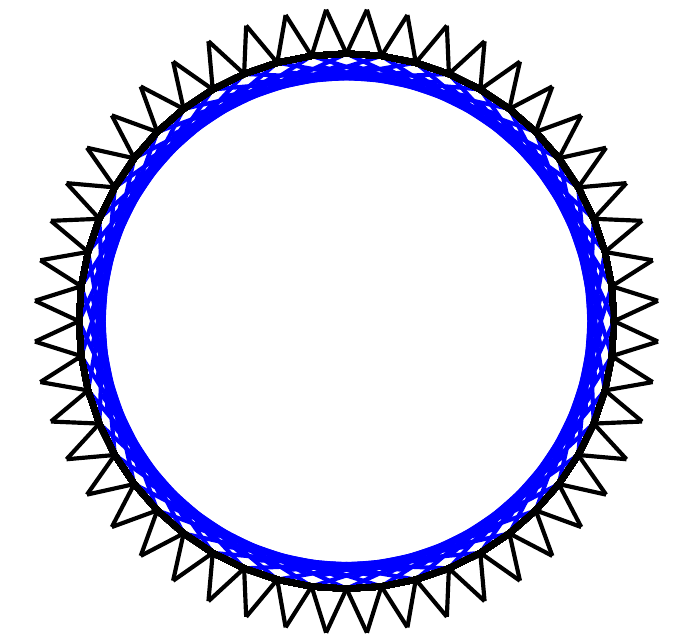}&\includegraphics[height=4cm]{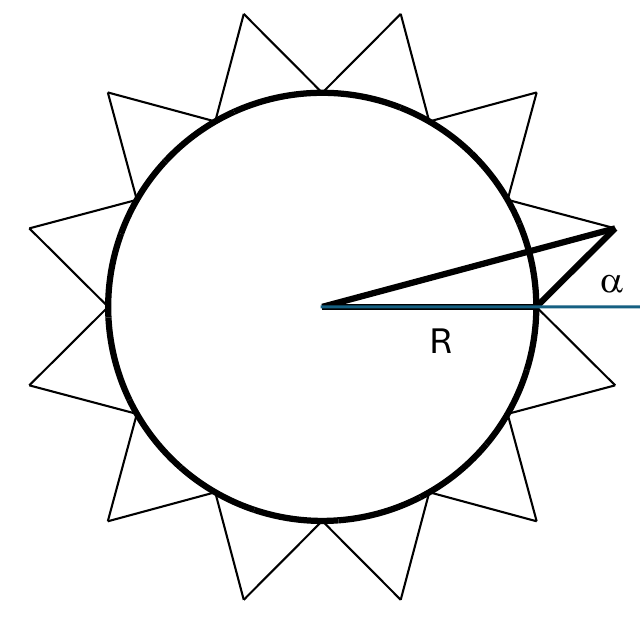}
	\end{tabular}
	\end{center}
   \caption[example] 
   { \label{Boivin} 
Illustration of the  Boivin radius  $R \cos(\alpha)$ as the envelop of rays perpendicular  to the edges of the teeth. The angle $\alpha$  must be as small as possible so that the Boivin radius approaches that of the central disk, which requires a very large number of short teeth not to mask the inner corona.}
   \end{figure} 

The idea that leads to serrated external occulters for solar astronomy is of geometric origin. It is assumed  that wave diffraction occurs over planes  perpendicular to the edges of the screen. Triangular teeth are placed on the edge of the circular occulters to move light away from the center of the optical axis.  Boivin \cite{boivin78} shown that  a dark disk of radius $R \cos(\alpha)$ is sheltered from the rays diffracted by the edges, where $R$ is the occulter radius and $\alpha$ is the exterior angle of the half-teeth triangle as shown in the right drawing of Fig.\ref{Boivin}.  This effect was recently numerically verified  using the Maggi-Rubinowicz boundary integral \cite{Rougeot2018}.  Serrated occulters are a simplified version of petal occulters. A gain of several orders of magnitude in rejection could be obtained if the triangular teeth would be replaced by a petal having a more studied shape, for example a Sonine function \cite{Aime2013}, but no experiment of this type is currently envisaged. 

\begin{figure} [ht]
   \begin{center}
   \begin{tabular}{cc} 
   \includegraphics[height=3cm]{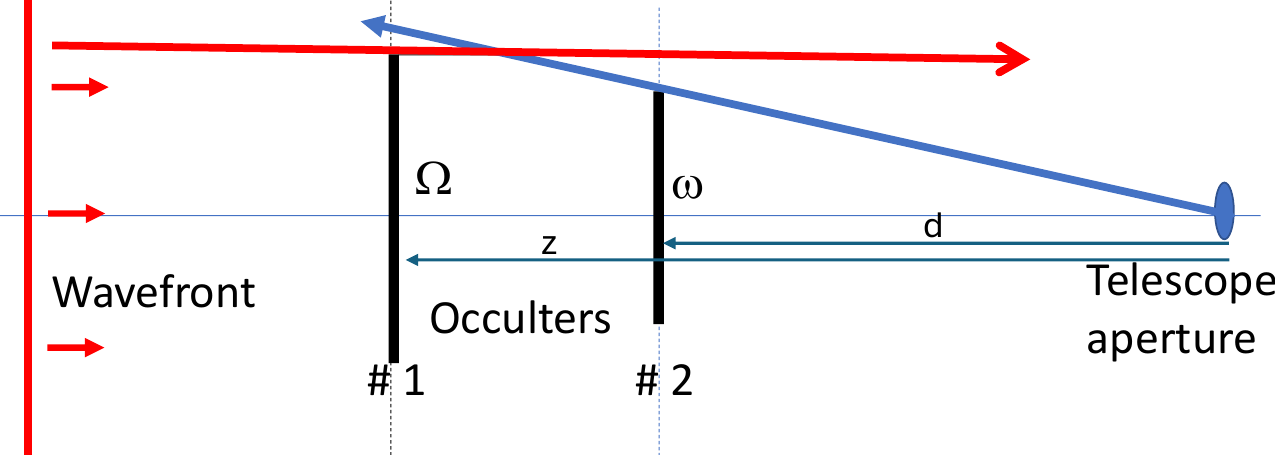} &
   \includegraphics[height=3.5cm]{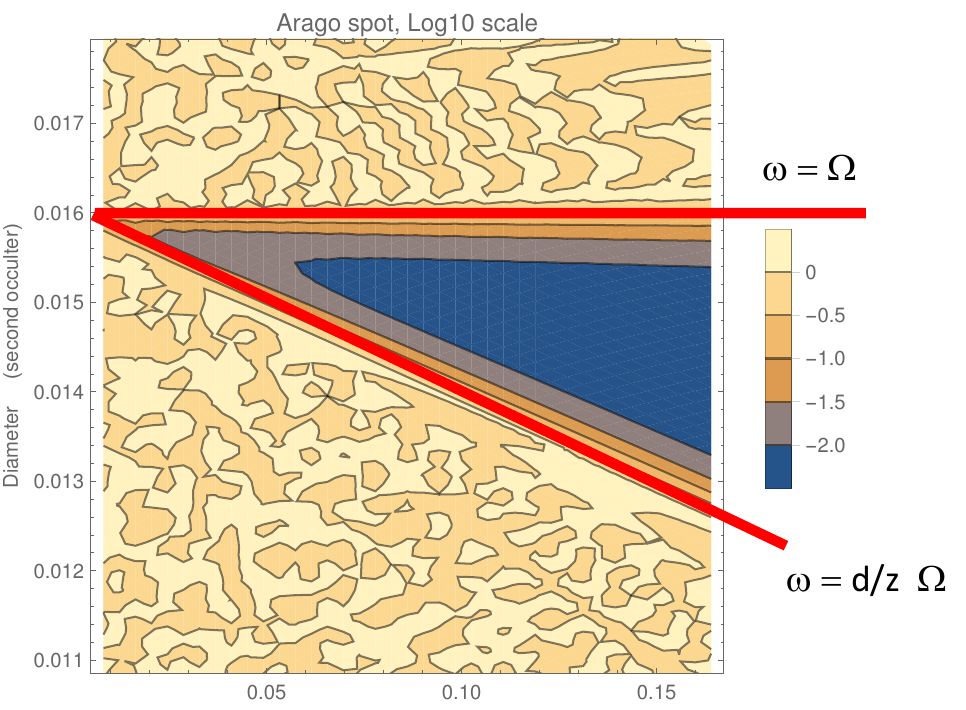}
	\end{tabular}
	\end{center}
   \caption[example] 
   { \label{MultipleDisc} 
Left, principle diagram: two disks $\#1$ and $\#2$ of diameters $\Omega$ and $\omega$ are at distances $z$ and $d$ from the entrance aperture of the telescope. Right: a substantial light reduction is obtained provided that  $\omega$ is half-way between the conditions  $\omega \le \Omega$ and $\omega \ge d/z \Omega$.}
   \end{figure}

The use of several disk occulters followed another reasoning. The idea is that the Arago spot exists because the edge of the circular occulter seen from the telescope is bright. The solution found by solar astronomers of the sixteen  is to darken this edge. Tousay {\cite{Tousey1965} } 
considered  that the best arrangement  was a system of several discs, so that each disk casts a shadow on the edge of the next disk. Fig.\ref{MultipleDisc} gives an illustration for 2 discs at the distances $z$ and $d$ of the entrance aperture. The second disc of diameter $\omega$ must be in the shadow of the first disc of diameter $\Omega$,  with  the mandatory requirement that from the telescope the first disc is not visible, so that $\Omega>\omega>d/z \Omega$. 
Analytical expressions exist for 2 and 3 disks but are difficult to evaluate numerically. We can generalize these calculations to more disks and obtain the ideal shape for a multi-disk occulter by continuity \cite{aime2020fresnel}. The exact calculation of the Fresnel diffraction of a true 3D occulter has not been carried out, to the limit of  knowledge of the authors of this paper.


 \begin{figure} [ht]
     \begin{center}
     \begin{tabular}{cc} 
\includegraphics[height=4.8cm]{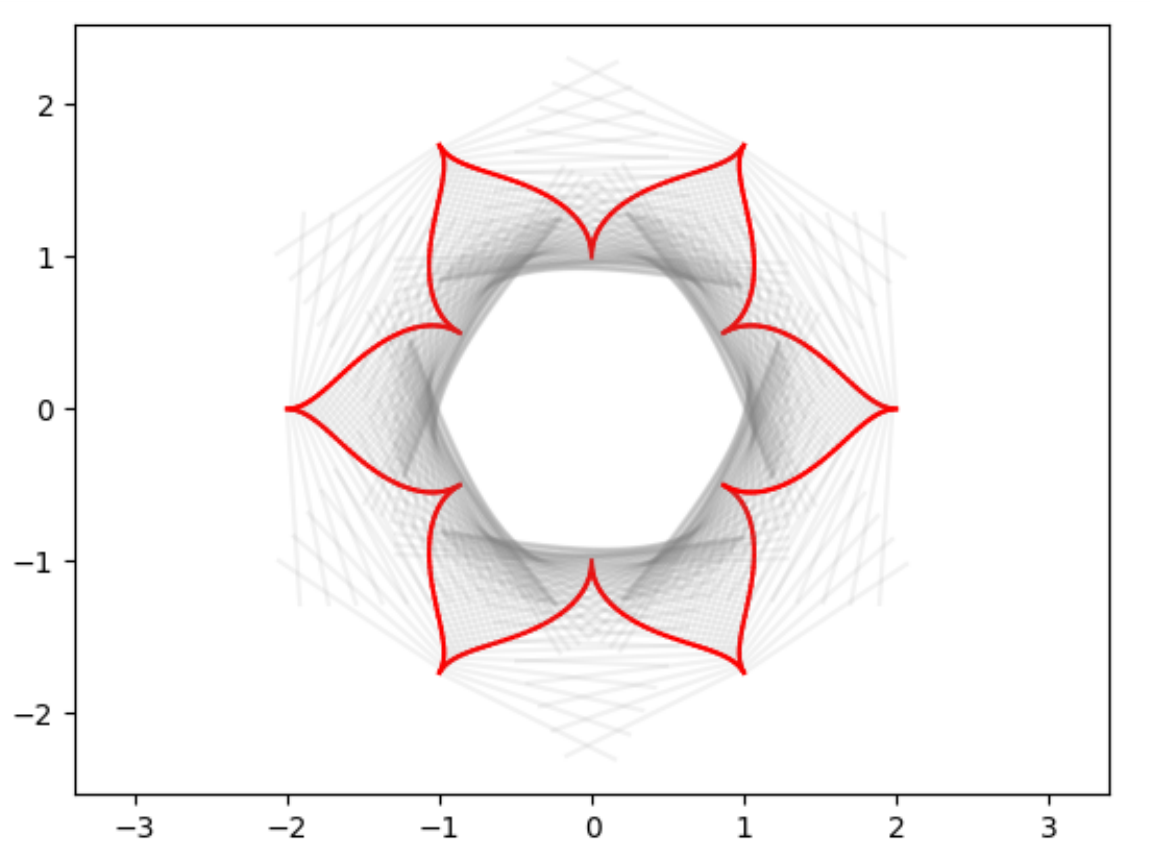} & \includegraphics[height=5cm]{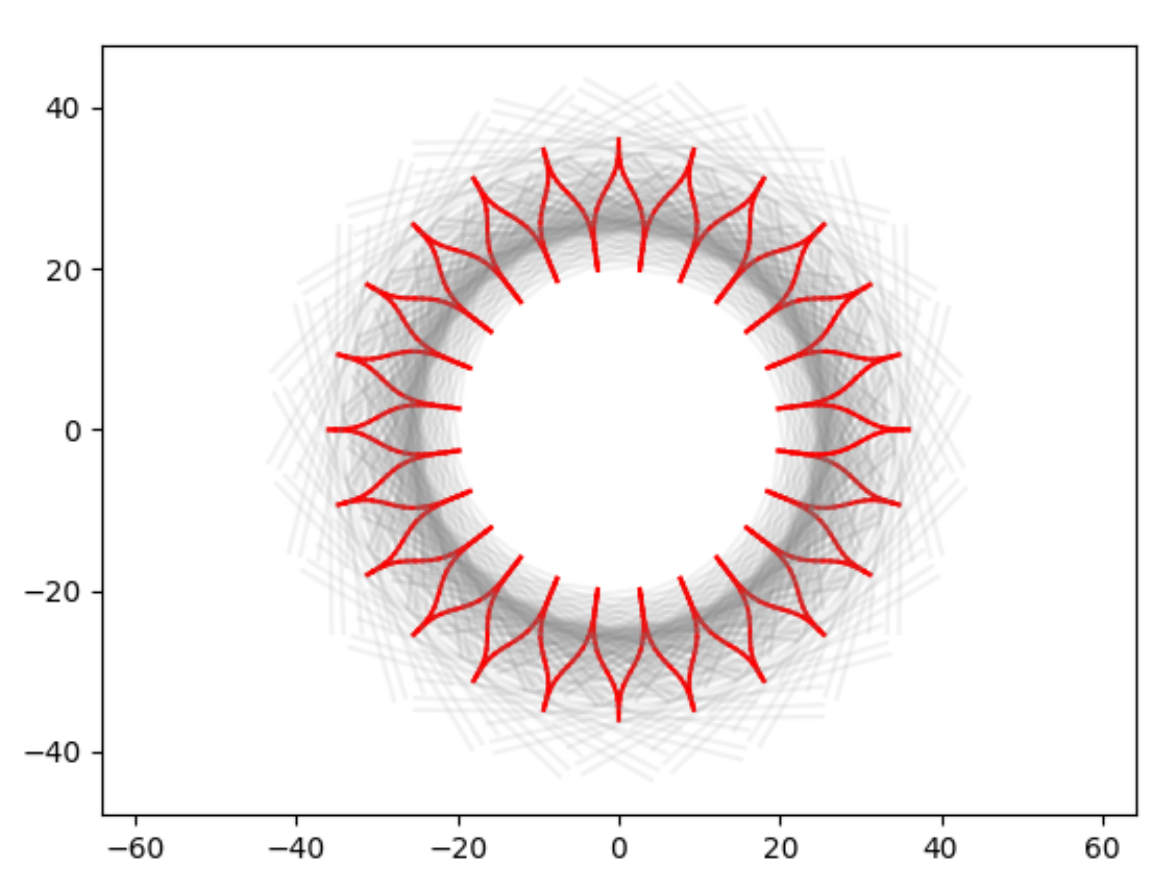}
\end{tabular}
\end{center}
    \caption{Illustration with the approach of Boivin using rays diffracting perpendicular to the edges of the petals for an occulter with 6 petals for which the profile is a mere cosine arch (left) and an occulter with 24 petals (right) corresponding to NW2 of SISTER which gives off a larger circular dark central area.}
    \label{BoivinStarshade}
\end{figure}
\section{Stellar external occulters (Starshade) }
\label{Starshade}
For stellar external occulters, the prevailing philosophy is to define an optimal apodized transmission \cite{cady2012} which makes it possible to obtain a very dark central zone  with an intensity less than $10^{-10}$ allowing to see   exoplanets \cite{Cash2006}, using the smallest possible shaped-occulter \cite{vanderbei2007},\cite{kasdin2009}, \cite{Flamary2014}. Considering that  these immense occulters cannot be manufactured  with variable transmission, shaped occulters \cite{marchal1985} are  used as the alternative to produce the central black zone. Figure \ref{BoivinStarshade} shows an illustration  of Boivin's empirical approach to these petal occulters. Here also as for the solar serrated occulter case, a dark central zone sheltered from the rays diffracting perpendicular to the edges of the petals is evidenced. A 24-petal occulter with a very studied profile is clearly more effective than a 6-petal occulter with a simpler profile.

 \begin{figure} [ht]
     \begin{center}
     \begin{tabular}{ccc} 
     \includegraphics[height=4.2cm]{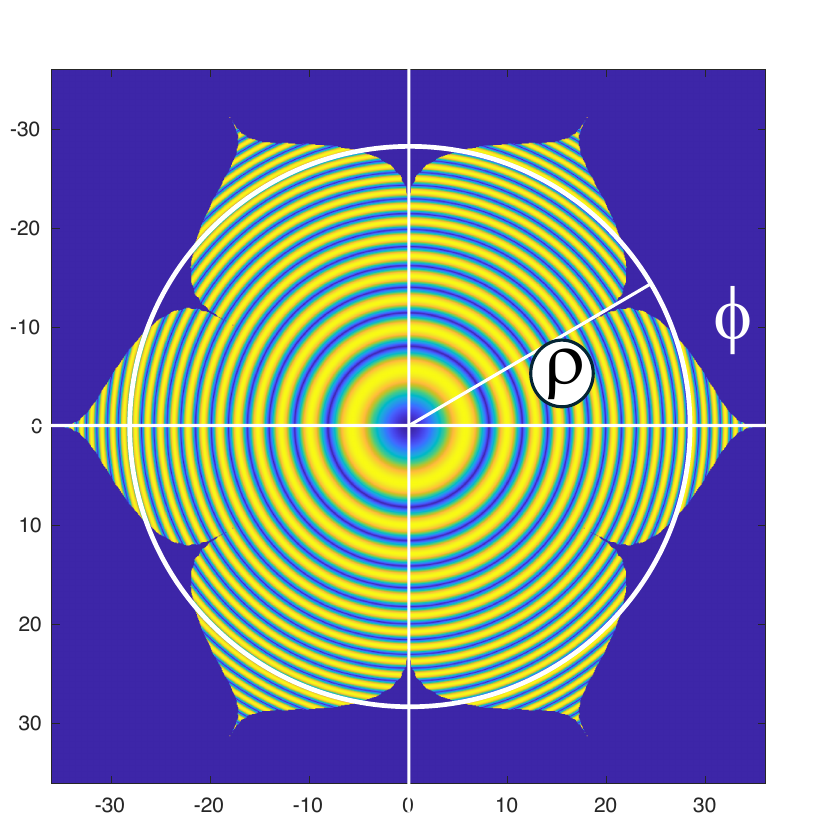}& \includegraphics[height=4cm]{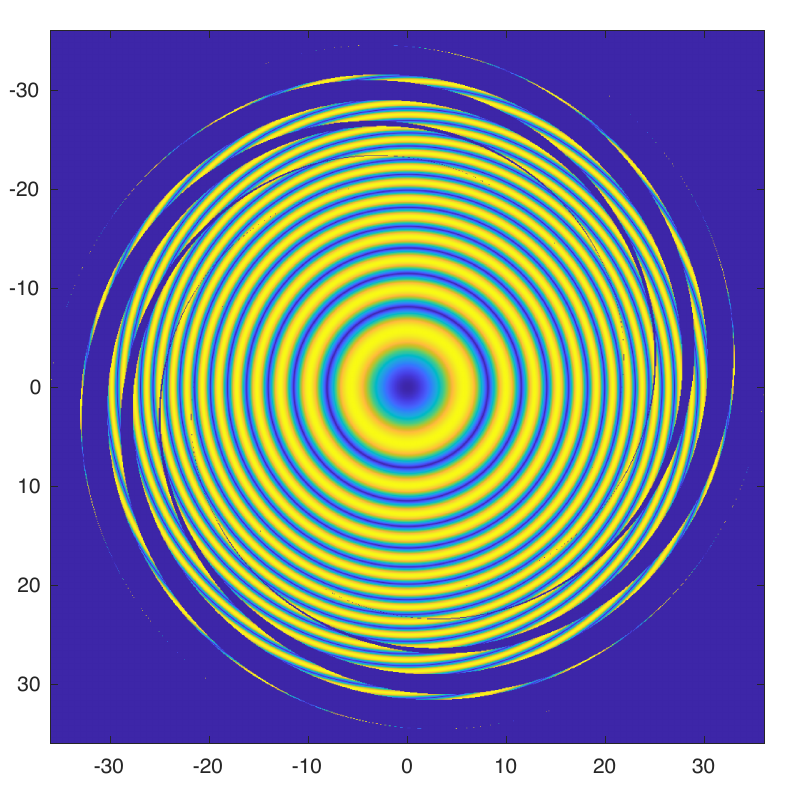} 
     & \includegraphics[height=4cm]{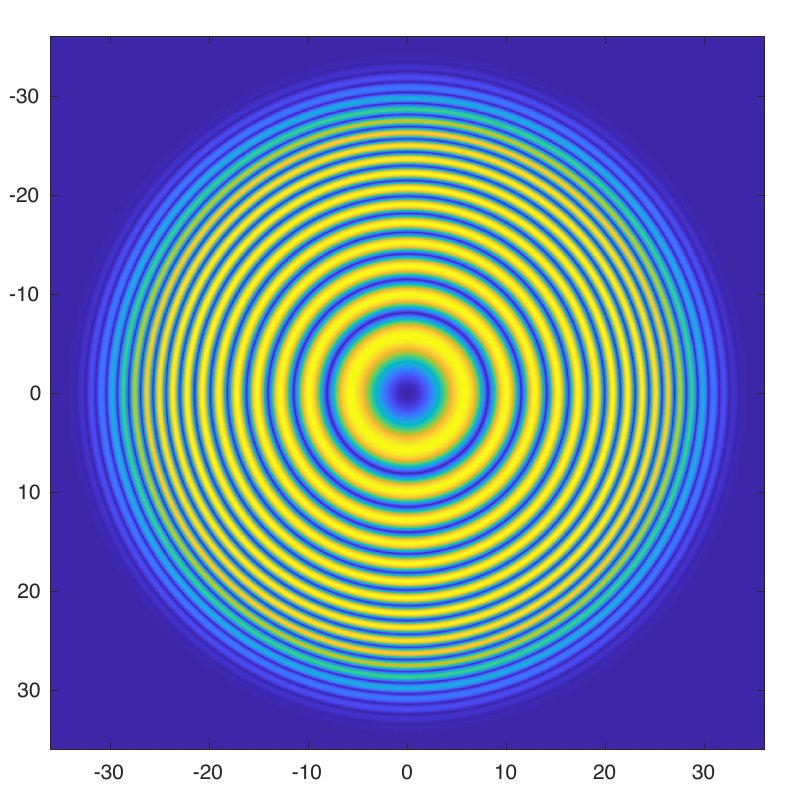} 
     \\
     \includegraphics[height=4.2cm]{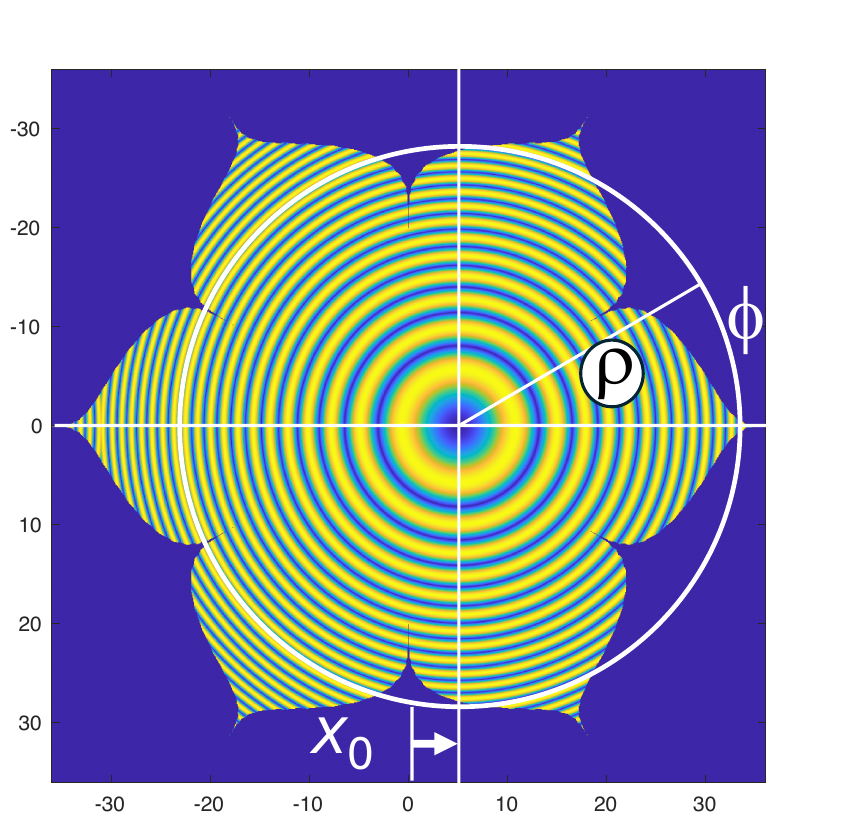}&\includegraphics[height=4cm]{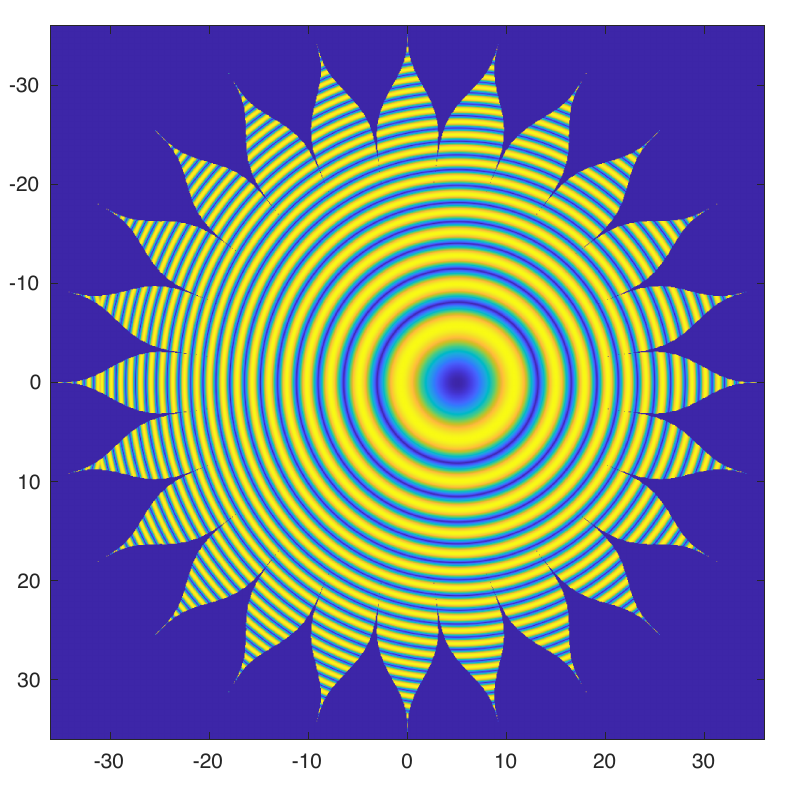} & \includegraphics[height=4cm]{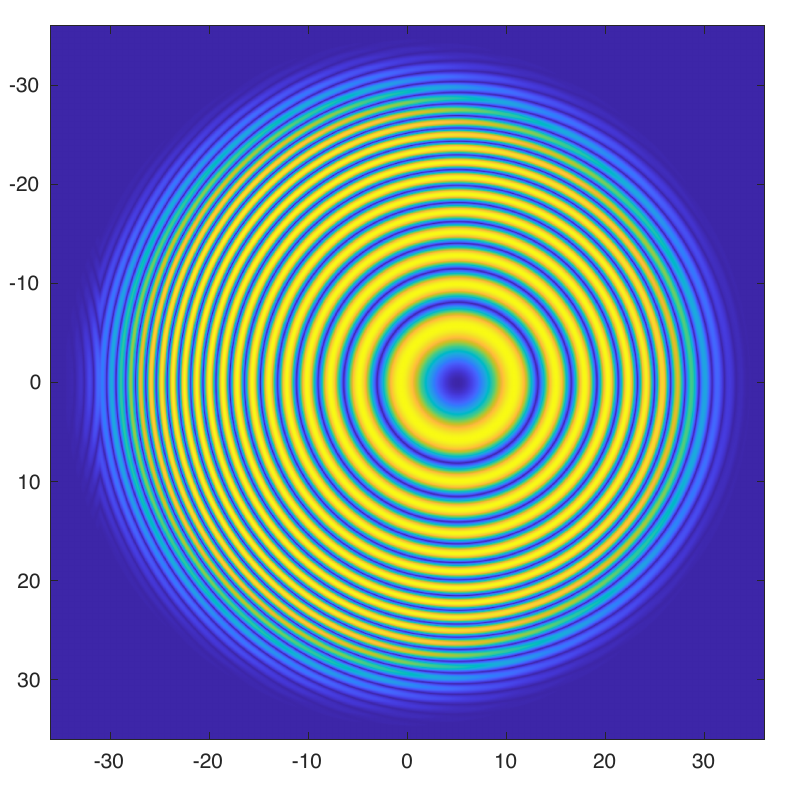}
     \end{tabular}
  	\end{center}
    \caption[example] { \label{IllustCircInt} 
  The function $g(\rho,X_0)$ of Eq.\ref{eq:g} corresponds to  the mean value along concentric circles centered at $x=X_0$ of the occulter transmission. In these figures, an illustration  of the quadratic phase of Eq.\ref{eq:conv}  is given behind. Top  panels are for $X_0=0$. The two left  occulters correspond to  the same value  $g(\rho,0)=p(\rho)$ that the apodized occulter at right, and therefore the same complex amplitude  $\Psi(0,0)$ at the center of the Fresnel diffraction pattern.    Bottom panels are for $X_0=5$ m, 6-petals and 24-petals occulters and the  apodized one. A drawing of $g(\rho,X_0)$ for different $X_0$ values is given in  Fig.\ref{StarshadePetals}.   }
    \end{figure} 



 \begin{figure} [ht]
    \begin{center}
    \begin{tabular}{ccc} 
    \includegraphics[height=4cm]{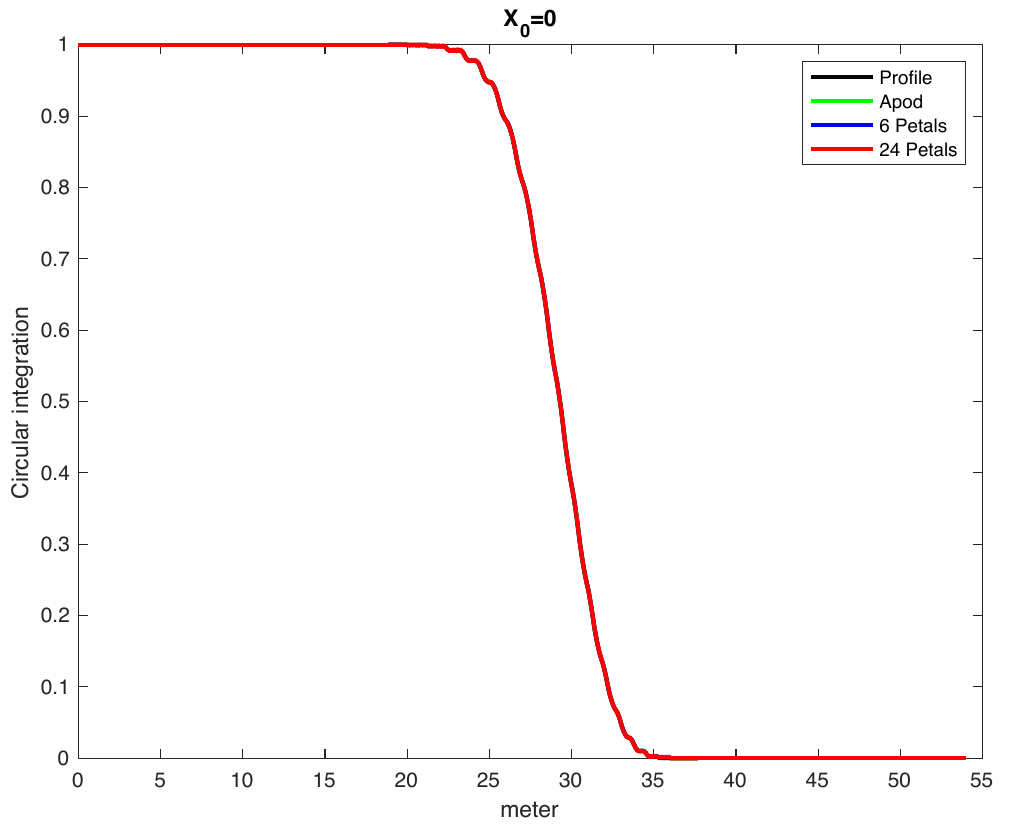} & \includegraphics[height=4cm]{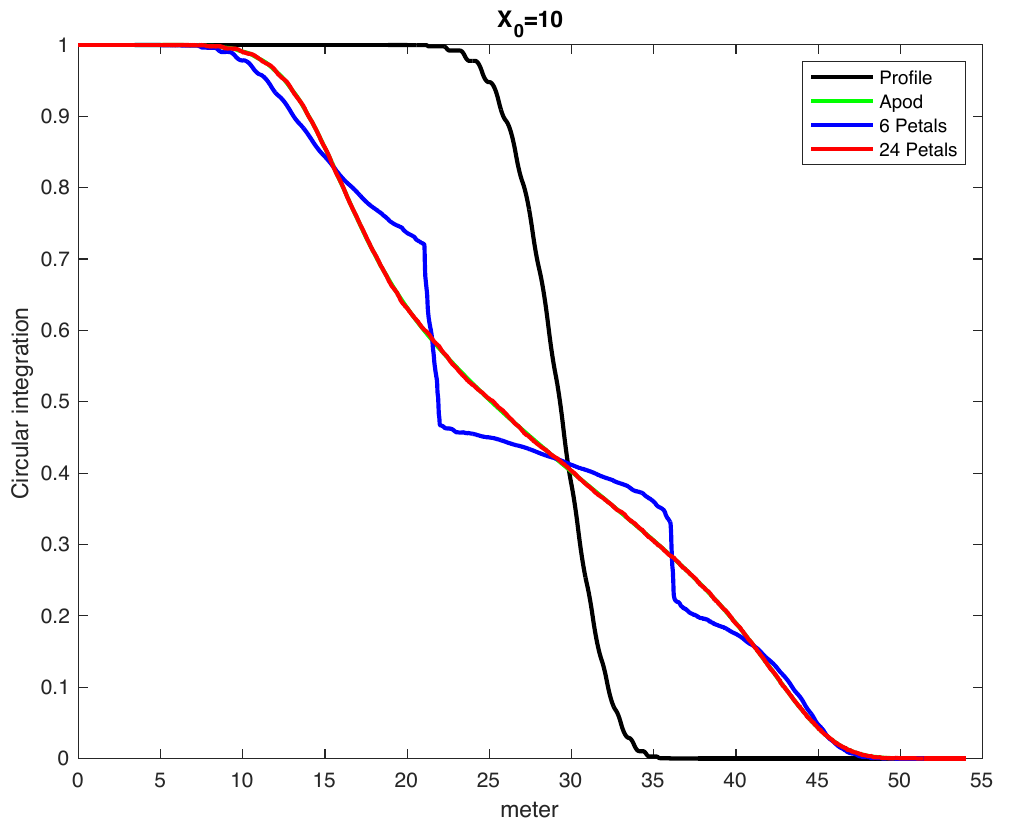} &
\includegraphics[height=4cm]{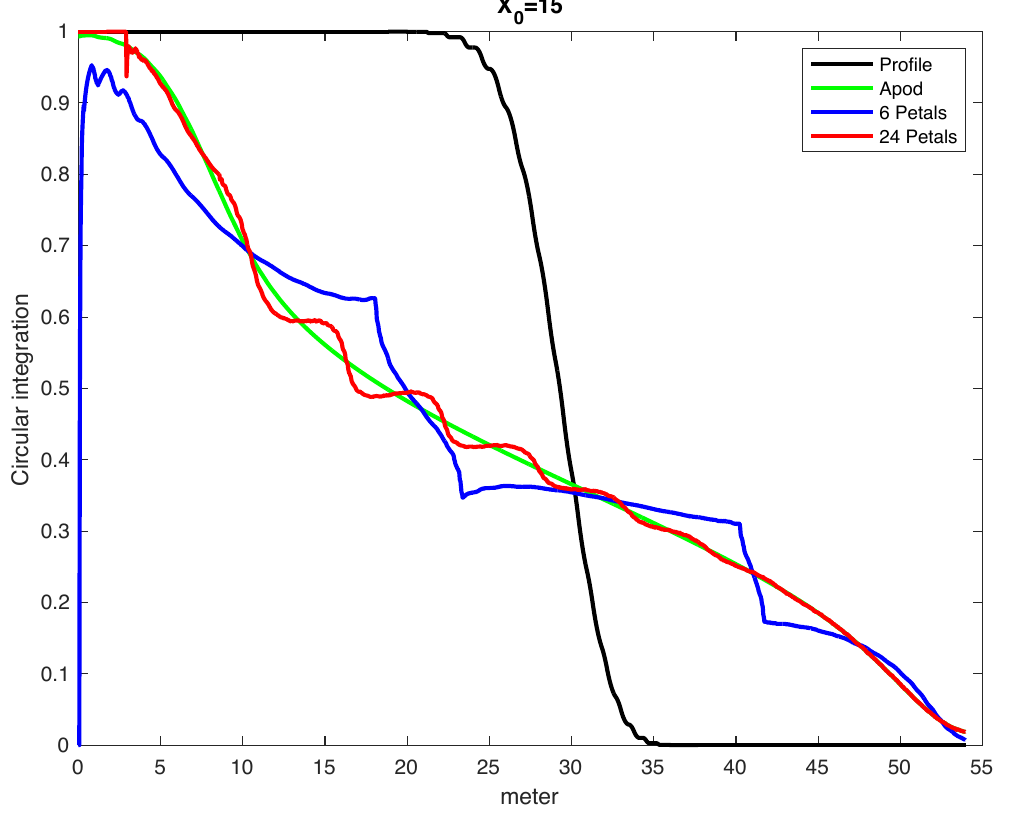}\\
   \includegraphics[height=4cm]{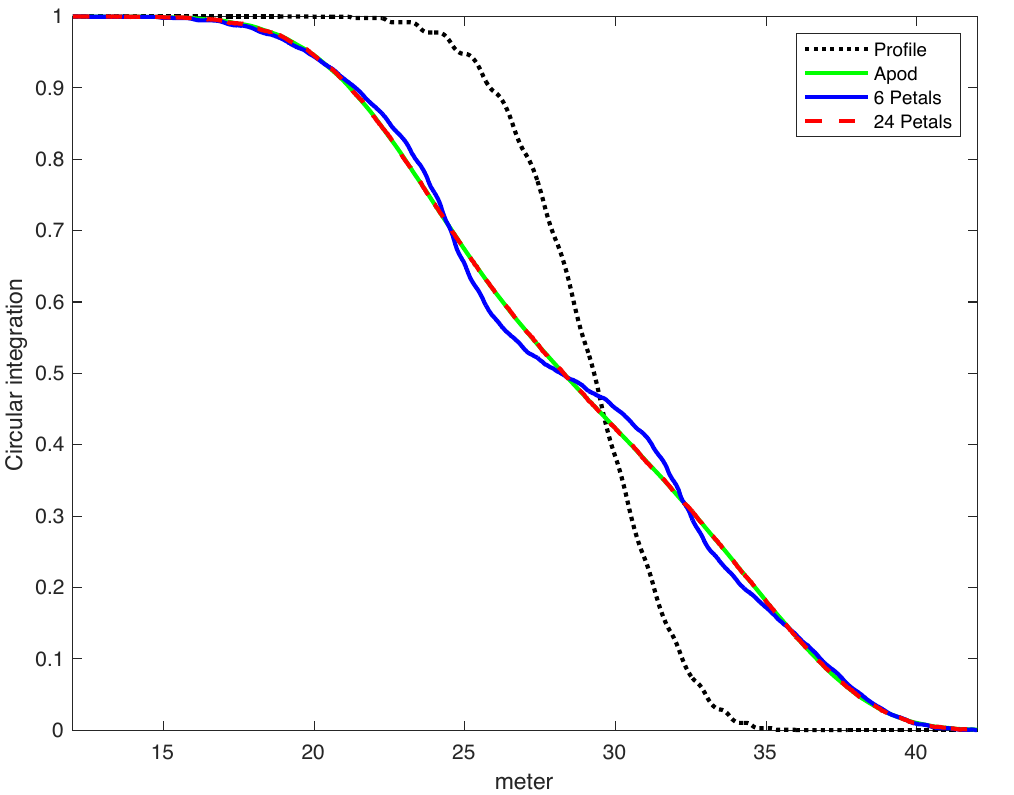}&\includegraphics[height=4cm]{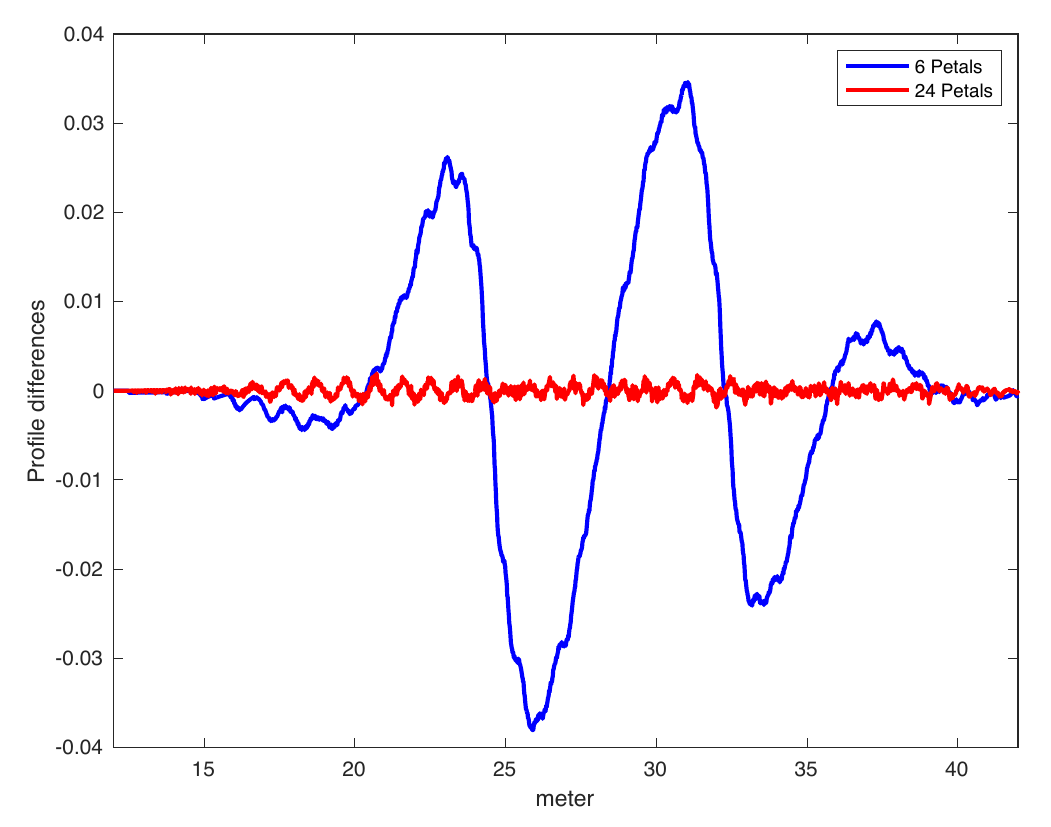} & \includegraphics[height=4cm]{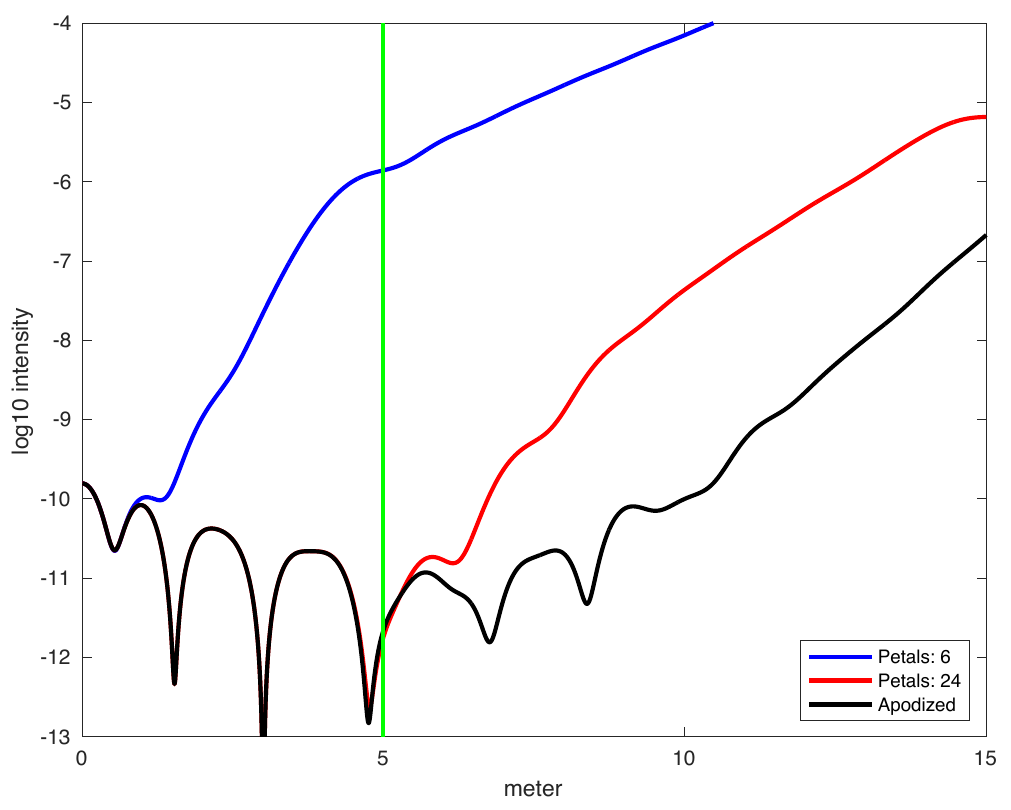} 
 	\end{tabular}
 	\end{center}
    \caption[example] 
    { \label{StarshadePetals} 
 Top panels, the functions  $g(\rho,X_0)$, for apodized,  6-petals and 24-petals occulters, for $X_0=$0, 10 and 15m. As expected all functions merge with  the profile $p(\rho)$ of NW2 of SISTER\cite{hildebrandt2021starshade} for $X_0=0$. Bottom panels correspond to the illustration $X_0=5$m given in Fig. \ref{IllustCircInt}. Left, the functions    $g(\rho,5)$, center, the differences $g_6(\rho,5)- g_{a}(\rho,5)$ and $g_{24}(\rho,5)- g_{a}(\rho,5)$, and right the intensity diffraction patterns (log10 scale) for the apodized, 6-petals and 24-petals occulters computed using the model of Vanderbei et al.\cite{vanderbei2007}. The vertical green line is at the position $X_0=5$m from center.   }
    \end{figure}

 For a more detailed analysis of the effects of the number of petals, consider the equation giving the Fresnel diffraction of an occulter $f(x,y)$ at the distance $z$ and at the wavelength $\lambda$ in the form of a convolution product:
\begin{equation}
\label{eq:general}
\Psi(x,y)=1-f(x,y)*\frac{1}{i \lambda z} \exp(i \pi \frac{x^2+y^2}{\lambda z}).
\end{equation}
For  $x=X_0,y=0$, this convolution can be written as:
\begin{equation}
\label{eq:conv}
\Psi(X_0,0)=1-\frac{1}{i \lambda z} \int \int\exp(i \pi \frac{\xi^2+\eta^2}{\lambda z})f(X_0-\xi,0-\eta) d\xi d\eta =1-\frac{2 \pi }{i \lambda z} \int g(\rho,X_0) \exp(i \pi \frac{\rho^2}{\lambda z})\rho^2 d\rho 
\end{equation}
where a change of Cartesian to radial coordinates was made to introduce the function
\begin{equation}
\label{eq:g}
 g(\rho,X_0)= \frac{1}{2 \pi \rho}\int_0^{2\pi} f(\rho \cos(\phi)+X_0,\rho \sin(\phi)) d\phi 
\end{equation}
that is the  integral of the occulter $f(x,y)$ along concentric circles centered at the point $(X_0, 0)$, taken as the new origin of the axes. An illustration of the integration is given in Fig.~\ref{IllustCircInt}. The result of the  integral is finally divided by $2 \pi \rho$ to obtain the average transmission on the  circles. 


For $X_0=0$, the petal-occulters are constructed such that  $g(\rho,0)=p(\rho)$, the optimal apodized transmission \cite{Cash2006}. In other words, the average of the all-or-nothing petal transmissions (1 or 0) on the concentric circles recovers the ideal variable transmission, here the profile NW2\cite{hildebrandt2021starshade}, which cannot be physically realized
 otherwise.
The three occulters of the top panels of Fig.~\ref{IllustCircInt}, the 6-petals occulter (left), a spiral version of it obtained using a twist of the petals (middle), complete the request and give the same mean circularly integrated value that the apodized occulter (right). 
Many shaped occulters may fit this requirement.

The $X_0 \neq 0$ case is illustrated in the bottom panels of Fig.~\ref{IllustCircInt} for $X_0 = 5$m. There we have considered the integrals for a 6-petals and a 24-petals occulter, and for the apodized occulter of NW2. The circular averages are then all different.

An illustration of $g(\rho,X_0)$  for $X_0 = 0, 5, 10$ and 15 m is given in Fig.\ref{StarshadePetals} for the 6 and 24 shaped occulters $g_6(\rho,5), g_{24}(\rho,5)$  and the apodized  occulter $g_a(\rho,5)$. As a reference  $p(\rho)$ is drawn in these figures. The function 
$g(\rho,X_0)$  departs  from $p(\rho)$, and spread out as $X_0$ increases.  The calculations are made in the direction of a tip of the petals, the curves are different in the direction of a hollow, but without changing the conclusion for the comparison between occulters. It appears in any case that $ g_{24}(\rho,X_0)$ is much closer to $g_a(\rho,X_0)$ than $g_6(\rho,X_0)$, and so are the diffraction patterns as expected by the theory of Vanderbei et al.\cite{vanderbei2007}.  
This is well evidenced by  differences between curves.  
We can consider a priory that the best shaped occulter is the one whose function $g(\rho,X_0)$ is as close as possible to $g_a(\rho,X_0)$. 
\begin{figure}  [ht]
    \centering
    \includegraphics[height=6cm]{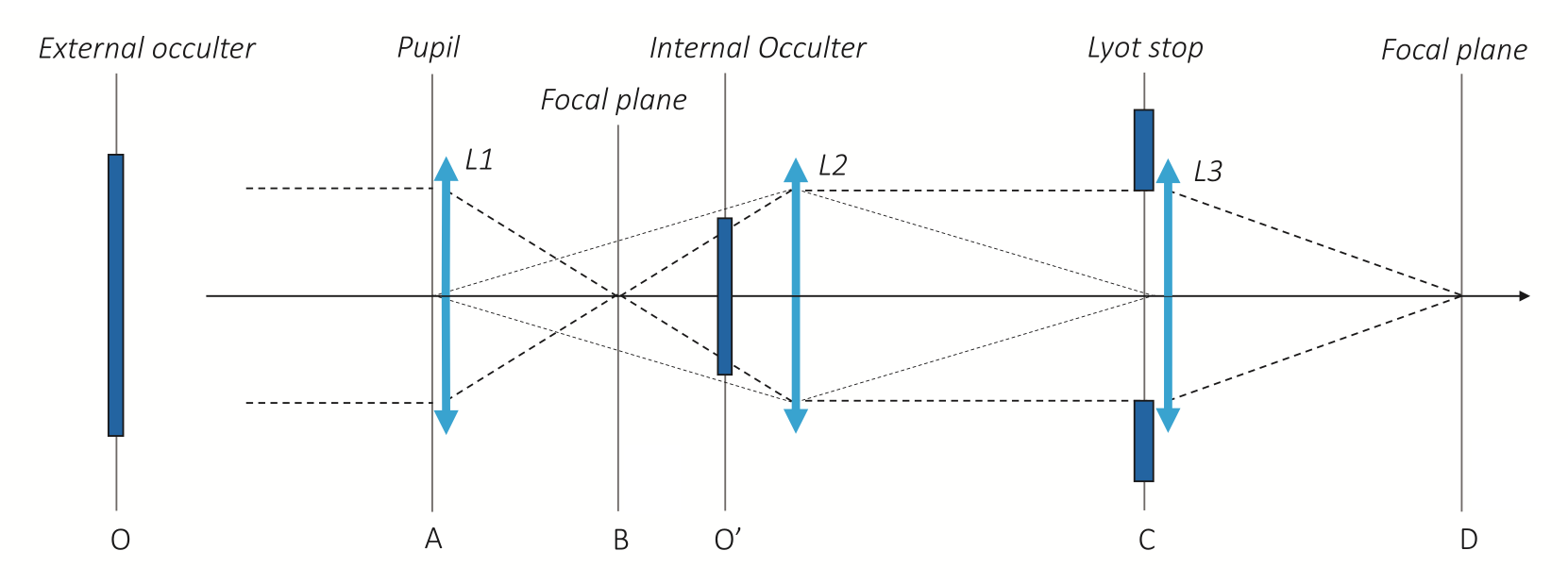} 
    \caption{Schematic diagram of a Lyot coronagraph coupled to an external occulter. Planes A, B, C, D are the classic planes of the Lyot coronagraph. The Lyot mask is in plane B, the Lyot stop in plane D. The presence of the external occulter (O) modifies the device such that the internal occulter must be optimally placed on the image (O') of the occulter. For exoplanets, we cannot differentiate O' from B.}
    \label{SchemaLyotShade}
\end{figure}
\vspace*{0.2cm}
\begin{figure}  [ht]
    \centering
    \includegraphics[height=7cm]{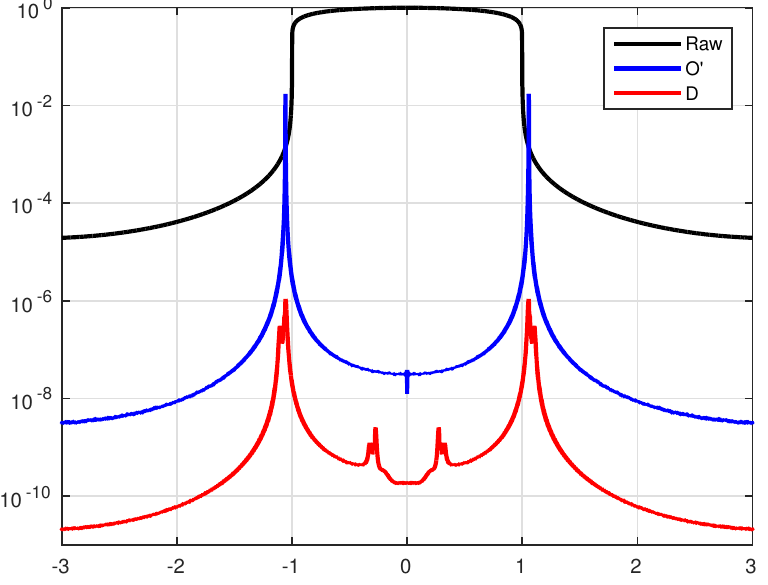}
    \caption{ Illustration of the effect of the Lyot coronagraph on the rejection of stray light for the ASPIICS solar coronagraph\cite{roug17}.  The black curve is the direct image of the solar photosphere (plane B) with a straylight at $10^{-4}$,  the blue curve gives the effect of the external occulter alone (plane O'), straylight at $10^{-8}$, the red curve the effect of the complete system (plane D), straylight at $10^{-10}$. The internal mask has a central hole for satellites positioning purposes. }
    \label{SolarResult}
\end{figure}

\begin{figure}  [ht]
    \centering
    \begin{tabular}{cc} 
    \includegraphics[height=6cm]{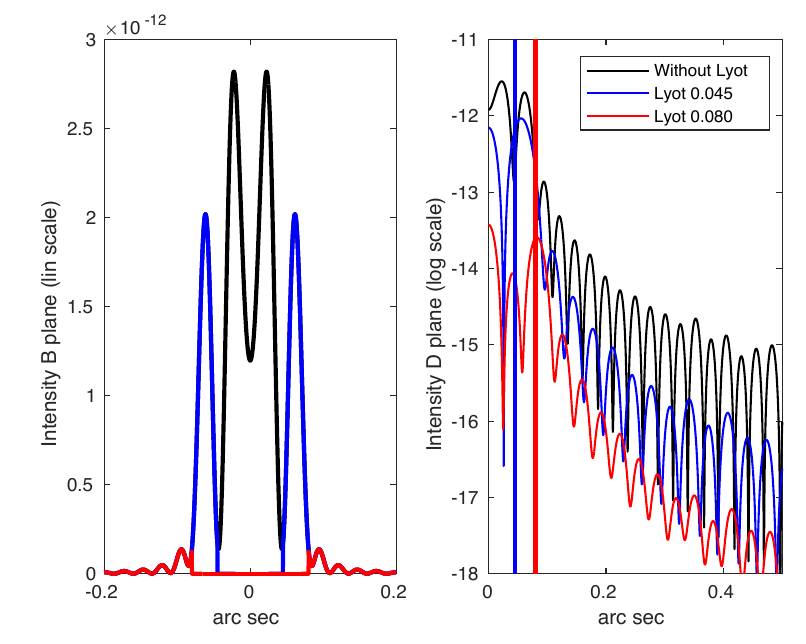} &\includegraphics[height=3.5cm]{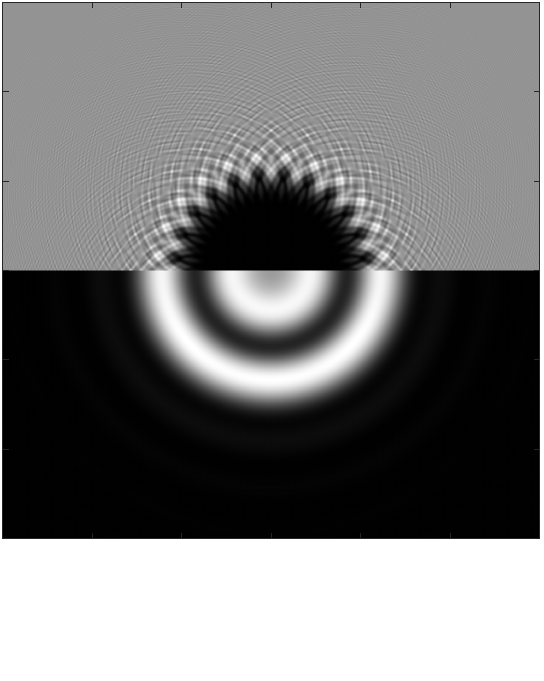} 
    \end{tabular}
    \caption{Left: residual image of the star (plane O', in black) with two Lyot masks removing the first ring (in blue) or the second ring (in red).  Middle: corresponding effects in the final plane D using a severe Lyot stop (diameter reduction 0.8). Only the largest mask allows notable rejection but at the expenses  of image reconstruction in the vignetting zone. The vertical lines give the diameters of the Lyot masks. The figure on the right is a hybrid 2D representation that shows the two straylight rings and occulter Fresnel diffraction projected onto the sky. }
    \label{Stellar}
\end{figure}

\begin{figure}  [ht]
    \centering
    \includegraphics[height=8cm]{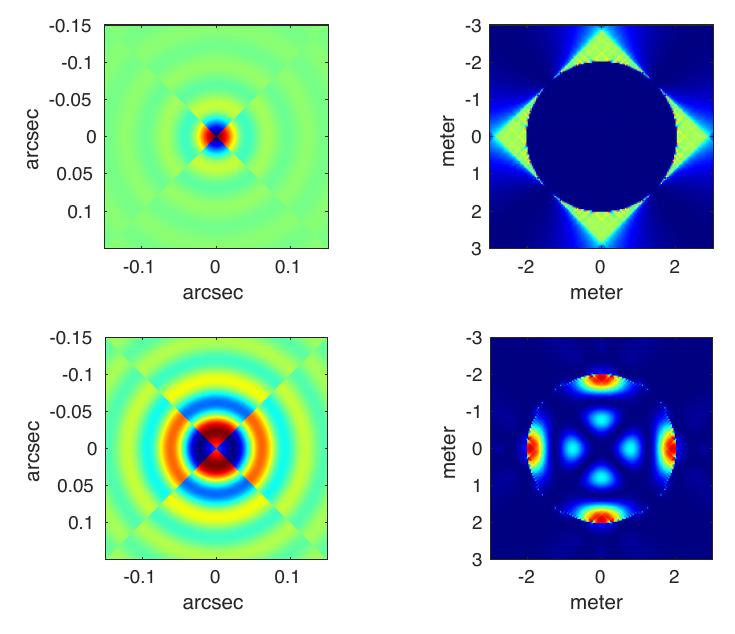}
    \caption{Illustration of the lack of efficiency of a 4-quadrant system with a Starshade. The top left figure corresponds to the Airy pattern (real part of the amplitude) given by a circular aperture   with the $\pm 1$ transmission  of the 4 quadrant mask. The top right figure is the corresponding  intensity in 
  plane C. A total rejection of the starlight in plane D can be obtained using a Lyot stop in plane C (see Fig.\ref{SchemaLyotShade} for planes denomination).   Bottom figures are for the same planes  with  the external occulter. There is no more an Airy pattern in the plane B and the wave remains unfortunately inside the aperture image in plane C.  }
    \label{4QC}
\end{figure}

\section{Benefit of adding an internal coronagraph}
\label{Lyot} 

For the observation of the solar corona, a complementary Lyot coronagraph is used. Figure \ref{SchemaLyotShade} gives the schematic diagram of a Lyot coronagraph coupled to an external occulter.  The original Lyot coronagraph is slightly modified in order to set the mask (called the internal occulter) on the image of the external occulter. To simplify numerical calculations it is interesting to add a  converging lens in plane A so as to reject the occulter to infinity and to compensate it with a diverging lens of the same negative power in C. The propagation of the wave is then done  using three Fourier transforms, from A to O', O' to C, and C to D. Figure \ref{SolarResult} shows that the  coupling external occulter and Lyot coronagraph is very effective, and results in a stray light below $10^{-10}$, for  perfect optics.

For the detection of exoplanets, projects are either with an external occulter alone or with an internal coronagraph alone. The schematic diagram of Fig.\ref{SchemaLyotShade} would still describe a system using a Lyot coronagraph after the starshade. The same calculation procedure can be made. In this case, given the very large distance where the external occulter is located, it will no longer be possible to differentiate experimentally plane O' from plane B. We give in Fig.\ref{Stellar}  the effect of a Lyot coronagraph with a small  mask of size the first ring, and with a large mask of size the second ring, using a very strong Lyot stop. The small mask will not affect the observation of exoplanets, but its effect on straylight is weak. The large mask is more effective in terms of straylight but it prevents the detection of exoplanets in the vignetting zone corresponding to  the petals edges.

Finally, one might wonder whether a phase filter system would be more efficient than a Lyot coronagraph. Fig.\ref{4QC} illustrates that a 4 quadrant phase mask placed at the focus of a telescope \cite{rouan2000} is not efficient either.

 

\bibliography{report2} 
\bibliographystyle{spiebib} 

\end{document}